\documentclass{bmcart}

\usepackage[utf8]{inputenc} 

\def\includegraphics{}

\startlocaldefs
\endlocaldefs

\usepackage{moreverb,url}
\usepackage{bm} 
\usepackage{bbm} 
\usepackage{soul}
\usepackage{mathrsfs,amsmath}
\usepackage{enumitem} 
\usepackage{varwidth} 
\usepackage{booktabs}

\newcommand{\R}{{\normalfont\textsf{R}}{}}
\newcommand{\code}[1]{\texttt{#1}}
\newcommand{\cond}{{\, \vert \,}}

\begin{document}

\begin{frontmatter}

\begin{fmbox}
\dochead{Research}


\title{A flexible approach for variable selection in large-scale healthcare database studies with missing covariate and outcome data}


\author[
   addressref={aff1}                   
]{\inits{JL}\fnm{Jung-Yi Joyce} \snm{Lin}}

\author[
   addressref={aff2},
   corref={aff2},
   email={liangyuan.hu@rutgers.edu}
]{\inits{LH}\fnm{Liangyuan} \snm{Hu}}

\author[
  addressref={aff3}                   
]{\inits{CH}\fnm{Chuyue} \snm{Huang}}

\author[
  addressref={aff2}                 
]{\inits{JJ}\fnm{Ji} \snm{Jiayi}}

\author[
  addressref={aff1}                 
]{\inits{SL}\fnm{Steven} \snm{Lawrence}}

\author[
  addressref={aff1}                  
]{\inits{UG}\fnm{Usha} \snm{Govindarajulu}}


\address[id=aff1]{
  \orgname{Department of Population Health Science and Policy, Icahn School of Medicine at Mount Sinai}, 
  \street{1425 Madison Ave},                     %
  \postcode{10029}                              
  \city{New York},                              
  \cny{USA}                                    
}

\address[id=aff2]{%
  \orgname{Department of Biostatistics and Epidemiology, Rutgers University},
  \street{683 Hoes Lane West},
  \postcode{08854}
  \city{Piscataway},
  \cny{USA}
}

\address[id=aff3]{
  \orgname{Primary Research Solution LLC.}, 
  \street{115 W 18th St},                     %
  \postcode{10011}                                
  \city{New York},                              
  \cny{USA}                                    
}



\end{fmbox}


\begin{abstractbox}

\begin{abstract} 
\parttitle{Background} 
Prior work has shown that combining bootstrap imputation with tree-based machine learning variable selection methods can provide good performances achievable on fully observed data when covariate and outcome data are missing at random (MAR).  This approach however is computationally expensive, especially  on large-scale datasets. 

\parttitle{Methods} 
 We propose an inference-based method, called RR-BART, which leverages the likelihood-based Bayesian machine learning technique, Bayesian additive regression trees, and uses Rubin's rule to combine the estimates and variances of the variable importance measures on multiply imputed datasets for variable selection in the presence of MAR data. We conduct a representative simulation study to investigate the practical operating characteristics of RR-BART, and compare it with the bootstrap imputation based methods. We further demonstrate the methods via a case study of risk factors for 3-year incidence of metabolic syndrome among middle-aged women using data from the Study of Women's Health Across the Nation (SWAN).

\parttitle{Results} The simulation study suggests that even in complex conditions of nonlinearity and nonadditivity with a large percentage of missingness,  RR-BART can reasonably recover both prediction and variable selection performances,  achievable on the fully observed data. RR-BART provides the best performance that the bootstrap imputation based methods can achieve with the optimal selection threshold value. In addition, RR-BART demonstrates a substantially stronger ability of detecting discrete predictors. Furthermore, RR-BART offers substantial computational savings. When implemented on the SWAN data, RR-BART adds to the literature by selecting a set of predictors that had been less commonly identified as risk factors but had substantial biological justifications. 
 
\parttitle{Conclusion} The proposed variable selection method for MAR data, RR-BART, offers both computational efficiency and good operating characteristics and is utilitarian in large-scale healthcare database studies.   

\end{abstract}




\end{abstractbox}
%

\end{frontmatter}



\section*{Background}

The problem of variable selection is one of the most popular model selection problems in statistical applications \cite{george2000variable}. Variable selection involves modeling the relationships between an outcome variable and a set of potential explanatory variables or predictors and identifying a subset of predictors that has the most impact on the model fit. Variable selection in the presence of missing  data has gained growing attention in recent years, particularly among statistical practitioners working with health datasets, which frequently present the missing data issue. 

There are three general missing data mechanisms: missing completely at random (MCAR), missing at random (MAR), and missing not at random (MNAR) \cite{little2012prevention}. 
When the missingness depends neither on observed data nor on the missing data, the data are said to be MCAR.  When data are MCAR, the complete cases are a random subsample of the population, thus results from a complete cases analysis would not be biased but can be less efficient when a large amount of data are discarded. A more realistic assumption about the missing data is  that the mechanism of missingness may depend on the observed data, and then the missing data are MAR given the observed data \cite{sterne2009multiple}. Under the MAR assumption, one can impute the missing values based on the observed data. In a more challenging situation where the missingness depends on the missing data, the data are MNAR \cite{little2019statistical}.  To handle MNAR, an approach recommended by the National Research Council \cite{national2010prevention}  is sensitivity analysis \cite{hogan2014bayesian, hu2018modeling}, which evaluates the impact of the potential magnitude of departure from MAR on analysis results. Little et al. \cite{little2019statistical} provide a comprehensive review of existing statistical approaches for handling missing data. In this article, we focus on the MAR mechanism that allows replacing missing data with substituted values or imputation based on the observed data, which is widely accepted in epidemiological and health research. 

Under MAR, variable selection can be conducted in combination with  imputation for missing values \cite{long2015variable}.  This approach is conceptually straightforward and is widely applicable to general missing data patterns \cite{long2015variable}.  The key consideration is how to combine the uncertainties about the imputation and the selection of predictors.

Wood et al. \cite{wood2008should} compared strategies for combining  the backward stepwise selection approach and multiple imputation for incomplete data, and recommended selecting variables based on Rubin’s rules that combine estimates of parameters and standard errors 
across multiple imputed data. Long and Johnson \cite{long2015variable} proposed combining bootstrap imputation  and stability selection \cite{meinshausen2010stability}. 
Bootstrap imputation is also known as “BS-then-MI”. The term refers to bootstrapping first, and then applying the imputation method within each bootstrap sample.
In Long and Jonhson \cite{long2015variable}, single imputation is performed on each bootstrap sample, and then in stability selection, the randomized lasso is applied to each of $M$ bootstrap samples; a set of predictors are deemed as important  if they are selected in at least $\pi M$ samples, where $\pi$ is a proportion representing the selection threshold. 
Both studies rely on  the parametric assumptions, in which the exact  relationships between response and covariates need to be made explicit. 

Flexible nonparametric methods can mitigate the reliance on the parametric assumptions and improve the variable selection results \cite{bleich2014variable,mazumdar2020comparison,ungaro2020machine,hu2020tree,hu2020quantile,hu2020machine,hu2020ranking,ji2020identifying}. A recent work by Hu et al. \cite{hu2021variable} investigated a general strategy that combines bootstrap imputation with six variable selection methods (four tree-based and two parametric) for variable selection among incomplete data. Their numerical studies suggest that  flexible tree-based methods achieved substantially better variable selection results than parametric methods with MAR data in conditions of nonlinearity and nonadditivity. Among the four tree-based methods, permutation-based Bayesian additive regression trees (BART) \cite{bleich2014variable} and recursive elimination based extreme gradient boosting (XGBoost) \cite{chen2016xgboost} were the top performers.  We term these two methods as BI-BART and BI-XGB, against which we will benchmark our proposed method for variable selection in the presence of MAR data. Although BI-BART and BI-XGB have good performance, they are  computationally expensive as they require both bootstrap and iterative variable selection procedures on each bootstrap sample. As a result, they are not easily scalable to large-scale healthcare data. In addition, their performance depends on the value of the selection threshold  $\pi$.

In this paper, we propose a new and simple strategy that leverages the likelihood-based Bayesian machine learning technique, BART, for variable selection with MAR data. By properly using the posterior distributions of the variable importance measures from the BART models on multiple imputed datasets, our proposed approach offers significant computational savings while still delivering excellent performance that is on par with the best performance BI-BART and BI-XGB can achieve with the optimal $\pi$. Furthermore, in addition to the commonly used performance metrics for variable selection \cite{wood2008should,bleich2014variable,hu2021variable}, we propose a way in which each method's ability to select important variables with incomplete data can be judged on the basis of out-of-sample predictive performance via the cross-validated area under the curve (AUC). Finally, we apply our proposed methods to re-analyze the Study of Women's Health Across the Nation (SWAN) data \cite{hu2021variable} and to identify predictors of 3-year incidence of metabolic syndrome among middle-aged women.

\section*{Methods}
\subsection*{Proposed methods}~\label{sec:bart}
Our proposed method uses BART. BART is a tree-based Bayesian probability model \cite{chipman2010bart}, offering an additional advantage over the algorithm-based machine learning methods of providing proper representations of uncertainty intervals via the posterior \cite{hu2020estimation, hu2021estimation,hu2021estimating}.  A regression or classification tree  approximates the covariate‐outcome relationship by recursive binary partitioning of the predictor space. The tree consists of the tree structure and all the decision rules sending a variable either left or right and leading down to a bottom node. Each of the bottom nodes represents the mean response of the observations falling in that node \cite{hu2020tree}.  For a binary outcome, BART uses probit regression and a sum of trees models, 
 \begin{eqnarray}
\mathrm{Pr}(Y= 1 \cond \bm{X}=\bm{x}) = \Phi\left(f(\bm{x})\right) = \Phi\left(\sum_{j=1}^m g(\bm{x}; \mathcal{T}_j, \mathcal{M}_j)\right),
\end{eqnarray}
where $\Phi (\cdot)$ is  the standard normal cumulative distribution function,  $\sum_{j=1}^m g(\bm{x}; \mathcal{T}_j, \mathcal{M}_j)$ is a sum of trees model, and $(\mathcal{T}_j, \mathcal{M}_j)$ are tree parameters. To avoid overfitting and limit the contribution of each $(\mathcal{T}_j, \mathcal{M}_j)$, a regularizing prior is put on the parameters, and the posterior is computed using Markov chain Monte Carlo (MCMC).  Details of the BART model can be found in Chipman et al. \cite{chipman2010bart}. 

 Here we leverage the underlying Bayesian probability model of the BART technique. We impute the missing data with random forest and use Rubin's rule \cite{rubin2004multiple,hu2019causal} to combine the estimates and variances of the variable inclusion proportion (VIP) -- the proportion of times each predictor is chosen as a splitting rule divided by the total number of splitting rules appearing in the model -- of each predictor provided by the BART model across multiply imputed datasets for variable selection. We refer to our proposed method as RR-BART, where ``RR'' stands for Rubin's rule. The key steps of RR-BART are: 
\begin{enumerate}
    \item 	Impute the data $M$ times using the $\code{missForest}$ technique \cite{stekhoven2012missforest}; fit a BART model to each of $M$ imputed datasets and draw $P$ Markov chain Monte Carlo (MCMC) posterior samples of the VIP for a predictor $X_k, \; k=1, \ldots, K$.  We then have a distribution of   $\text{VIP}_{kmp}$, $m=1, \ldots, M, \; p = 1, \ldots, P$.  \vspace{-1pt}
    \item Calculate the average VIP across $P$ posteriors and $M$ imputed datasets for each predictor $\overline{\text{VIP}}_{k\cdot\cdot}$, and identify the variable $X_{k'}$ with the minimum value, $\overline{\text{VIP}}_{k'\cdot\cdot} = \min\limits_{k=1,\ldots,K}(\overline{\text{VIP}}_{k\cdot\cdot})$. If $\overline{\text{VIP}}_{k'\cdot\cdot} > \frac{1}{2K}$, then stop the algorithm and no variable selection will be performed. Otherwise, calculate the distance between each VIP score and the minimum of average posterior mean VIPs,  $\Delta_{kmp} = \text{VIP}_{kmp} - \overline{\text{VIP}}_{k'\cdot\cdot}$ for $k=1, \ldots, K$, $m=1, \ldots, M$, $p=1, \ldots, P$.  \vspace{-1pt}
    \item Apply the Rubin's rule \cite{rubin2004multiple,hu2019causal} to the distribution of $\Delta_{kmp}$. The overall mean and total variance of the distance score for predictor $X_k$ are calculated as follows: 
    \begin{eqnarray*}
    \bar{Q}_k &= &\sum_{m,p} \Delta_{kmp} /MP, \\
   \text{Within-imputation variance } W_k&=&\frac{1}{M} \sum_m \text{Var}( \bar{\Delta}_{km\cdot} ), \\
  \text{Between-imputation variance } B_k &=& \frac{1}{M-1} \sum_m (\bar{\Delta}_{km.} - \bar{\Delta}_{k..})^2, \\
  \text{Total variance } T_k  &=& W_k + (1+\frac{1}{M})B_k,
    \end{eqnarray*}
 where $\text{Var}( \bar{\Delta}_{km\cdot} )$ is the variance among $\{\Delta_{kmp},p =1, \ldots, P\}$ divided by the sample size $n$, $\bar{\Delta}_{km.} = \frac{1}{P}\sum_p \Delta_{kmp}$, and $\bar{\Delta}_{k..} = \frac{1}{MP} \sum_{m}\sum_p \Delta_{kmp}$. 
 The $1-\alpha$  confidence interval for variable $k$'s average distance score is $\bar{Q}_k \pm t_{df, 1-\alpha} \sqrt {T_k}$, where the degrees of freedom for the $t$-distribution is $df = (M-1)/((B_k+B_k/M)/T_k)^2$ \cite{rubin2004multiple}. 

\item 	Select variables with the $1-\alpha$ confidence intervals that do not contain zero.
\end{enumerate}

The key idea of this approach is to properly characterize the probability 
distribution of the VIP measure for each predictor from multiple imputed datasets, and then identify the ``important'' predictors that have significantly larger VIPs. As pointed out by anonymous reviewers, this approach implies that there exists at least one irrelevant predictor so that variable selection is needed. In the extreme scenario where all candidate predictors are useful, the minimum of average posterior mean VIPs, $\overline{\text{VIP}}_{k'\cdot\cdot}$, will be considerably far away from zero, then variable selection will be deemed as unnecessary and thus all predictors will be selected (step 2).  In step 4, bigger values of $\alpha$ will lead to more selected variables and smaller $\alpha$ will result in less variables selected. We used $\alpha = 0.05$ as recommended in previous work \cite{kapelner2015prediction}. In step 1, we use the flexible imputation method $\code{missForest}$.  The $\code{missForest}$ proceeds as follows.  Initial guesses are made for the missing values (e.g., mean or mode), and variables are sorted in the ascending order of missingness proportions. The variables with missing data are in turn, in the descending order of missingness proportions, regressed via Random Forest on other variables. The missing values are imputed/updated using the fitted Random Forest model as the prediction model.  This process is repeated until a stopping criterion is met or the maximum number of iterations is reached. Detailed description of $\code{missForest}$ is provided in Supplementary Section 1 and in Stekhoven and B{\"u}hlmann  \cite{stekhoven2012missforest}. 

\subsection*{Comparison methods}
\paragraph{BI-BART} Hu et al. \cite{hu2021variable} proposed BI-BART,  which combines  bootstrap imputation with a permutation based variable selection method using BART, developed by Bleich et al.  \cite{bleich2014variable} The term ``BI'' refers to bootstrap imputation. The permutation based BART variable selection method uses the VIP.  The response vector is permuted a number of times (e.g, 100) and the BART model is fitted to each of the permuted response vectors and the original predictors; the VIPs computed from each of BART model fits constitute the null distributions. A variable is selected if its VIP from the BART model fitted to the unpermuted response is larger than the $1-\alpha$ quantile of its own null distribution.  The BI-BART method proceeds with the following three steps: (1) generate $B$ bootstrap data sets and impute missing data using $\code{missForest}$, (2) perform variable selection using BART on each bootstrap data set, (3) select the final set of variables if they are selected in at least $\pi B$ datasets, where $\pi$ is a fraction threshold between 0 and 1 for selecting a predictor.

\paragraph{BI-XGB} BI-XGB combines bootstrap imputation with XGBoost for variable selection among incomplete data.  At the core of the XGBoost method is gradient boosting \cite{chen2016xgboost}. Boosting is a process in which a weak learner is boosted into a strong learner.  The idea of gradient boosting is to build a model via additive functions that minimizes the loss function (exponential loss for classification), which measures the difference between the predicted and observed outcomes \cite{friedman2000additive}. XGBoost uses a gradient tree boosting model with each additive function being a decision tree mapping an observation to a terminal node. In addtion, XGBoost uses shrinkage and column subsampling to further prevent overfitting \cite{chen2016xgboost}. Shrinkage scales newly added weights by a factor after each step of tree boosting and the column subsampling selects a random subset of predictors for split in each step of tree boosting. Variable selection via XGBoost on fully observed data is carried out in a recursive feature elimination procedure \cite{hu2021variable}, using the variable importance score provided by the XGBoost model. Among incomplete data, BI-XGB carries out variable selection in three steps as described above for BI-BART, with permutation based BART method in step (2) replaced with XGBoost based variable selection method.

\paragraph{MIA-BART and MIA-XGB} As tree-based methods, both BART and XGBoost offer a technique,  missingness incorporated in attributes (MIA) \cite{kapelner2015prediction, chen2016xgboost}, to handle missing covariate data by treating the missingness as a value in its own right in the splitting rule.  The MIA algorithm chooses one of the following three rules for all variables for splitting $X$ and all splitting values $x_c$: (1) if $X$ is observed and $X\leq x_c$, send this observation left; otherwise, send this observation right. If $X$ is missing, send this observation left, (2) 
If $X$ is observed and $X\leq x_c$, send this observation left; otherwise, send this observation right. If $X$ is missing, send this observation right, (3) If $X$ is missing, send this observation left; if it is observed, regardless of its value, send this observation right. We refer to work by Kapelner and Bleich \cite{kapelner2015prediction} for a detailed description of the MIA procedure. As comparison methods, we  implement MIA within the BART model and within the XGBoost model while performing variable selection: MIA-BART and MIA-XGB. Because MIA cannot handle missing outcome data, we implement this technique on two versions of data:  (i) only cases with complete outcome data; (ii) data with imputed outcomes.  

\subsection*{Performance metrics} 

To stay consistent with the literature \cite{wood2008should,bleich2014variable, hu2021variable} so that our  methods and  results can be compared with previous work in similar contexts, we  assess the performance of variable selection using four metrics: (i) precision, the proportion of truly useful predictors among all selected predictors, (ii) recall, the proportion of truly useful variables selected among all useful variables, (iii) $F_1 =  \dfrac{2\text{ precision} \cdot \text{recall} }{\text{precision} + \text{recall}}$, the harmonic mean of precision and recall. The $F_1$ score balances between avoiding selecting irrelevant predictors (precision) and identifying the full set of useful predictors (recall), and (iv)  Type I error, the mean of the probabilities that a method will incorrectly select each of the noise predictors. Note that ``precision'' and ``positive predictive value (PPV)'' are sometimes used interchangeably; and the same goes for ``recall'' and ``sensitivity''.  Type I error is sometimes referred to as ``false positive'' in the machine learning literature. These performance metrics will be calculated among 250 replications for larger sample sizes $n=1000, 5000$ and among 1000 replications for smaller sample sizes $n=300, 650$.

In addition to the four metrics, we propose to use the cross-validated AUC to evaluate the prediction performance of each model with selected variables for incomplete data. This additional metric will be useful to 
distinguish between the methods in the situation where it remains unclear  which method is able to select the most relevant predictors based on the four metrics described above. The AUC evaluation is constituted of the following three steps: 
\begin{enumerate}
    \item [(1)] randomly split the data into two halves, and then implement each of the methods using half of the data to select variables, 
    \item [(2)] impute the other half of the data (single imputation) and record the AUC of each model with the selected variables. For BART based methods, calculate the AUC using BART, and for XGB based methods, calculate the AUC using XGBoost, and 
    \item [(3)] repeat steps (1) and (2) 100 times to get the distribution of the AUCs. 
\end{enumerate}

\subsection*{Simulation of incomplete data}
We adopt the simulation settings used in Hu et al. \cite{hu2021variable} to accurately mimic realistic missingness problems and to impartially compare methods in similar contexts. We use the multivariate amputation approach  \cite{schouten2018generating} to generate missing data scenarios with desired  missingness percentages and the missing data mechanisms. The multivariate amputation procedure first randomly divides the complete data into a certain number (can be one) of subsets, each allowing the specification of any missing data pattern. Then 
the weighted sum scores are calculated for individuals in each subset to amputate the data. For example,  the weighted sum score for $X_3$ and individual $i$ can take the form, $wss_{x_3,i} = x_{1i} w_1 + x_{2i} w_2$, which attaches a non-zero weight $w_1$ to $x_{1i}$ and $w_2$ to $x_{2i}$ to suggest that the missingness in $X_3$ depends on $X_1$ and $X_2$. A logistic distribution function \cite{van2018flexible} is then applied on the weighted sum scores to compute the missingness probability, which is used to determine whether the data point becomes missing or not. 

We consider simulation scenarios that represent the data structures commonly observed in health studies: (i) sample sizes ranging from small to large: $n= 300, 650, 1000, 5000$,  (ii) 10 useful predictors that are truly related to the responses, $X_1, \ldots, X_{10}$, and 10, 20, 40 noise predictors, where  $X_1$ and $X_2$ $\stackrel{i.i.d}{\sim}$ $\text{Bern}(0.5)$, $X_3$, $X_4$ and $X_5$ $\stackrel{i.i.d}{\sim}$  $N(0,1)$, $X_6$ $\stackrel{i.i.d}{\sim}$ $\text{Gamma}(4, 6)$, and $X_7, X_8, X_9, X_{10}$ were designed to have missing values under the MAR mechanism; a half of noise predictors are simulated from $N(0,1)$,  and the other half from $\text{Bern}(0.5)$, and (iii) 20\% missingness in $Y$ with 40\% overall missingness, and 40\% missingness in $Y$ with 60\% overall missingness. 
There are a total of 24 scenarios considered, including 4 sample sizes $\times$ 3 ratios of useful versus noise predictors $\times$ 2 missingness proportions. We additionally investigate how each method performs when there are no noise predictors for $n=1000$ and two missingness proportions. The outcomes are generated from a  model with arbitrary data complexity: (i) discrete predictors with strong ($x_1$) and moderate ($x_2$) associations; (ii) both linear and nonlinear forms of continuous predictors; (iii) nonadditive effects ($x_4x_9$). The outcome model is specified as: 
\begin{equation*}
\begin{split}
\mathrm{Pr}(y=1 \cond x_1,\ldots,x_{10}) = \text{logit}^{-1} \big(&-2.7+1.8x_1+0.5x_2+1.1x_3-0.4e^{x_5 }-0.4(x_6-3.5)^2\\
+0.3(x_7-1)^3+1.1x_8
&-1.1x_{10}+5\sin(0.1\pi x_4x_9)-0.4x_5 x_{10}^2+0.4 x_3^2 x_8\big).
\end{split}
\end{equation*}
Following generating the full data, the multivariate amputation approach was used to amputate  $X_7, X_8, X_9, X_{10}$ and $Y$ under the MAR mechanism to create the desired missingness characteristics. Detailed simulation set-up appears in Supplementary Section 2.

Because we generated four predictors, $X_7$, $X_8$, $X_9$ and $X_{10}$ from the normal distribution with the mean depending on other predictors, correlations among covariates $X_3, \ldots, X_{10}$ were established. The correlations range from  $-0.034$ to 0.413. Supplementary Table 1 shows the Pearson correlations among the predictors $X_3, \ldots, X_{10}$. An anonymous reviewer pointed out that we can measure the strength of MAR by first taking a missing indicator variable for each variable with missing values and regressing it on the other variables related to it being missing, and then estimating the AUC of the regression model. By this approach, the strength of missingness in $X_7, X_8, X_9, X_{10}$ and $Y$ are strong MAR with the AUC ranging from .72 to .92; see Supplementary Table 2.

\subsection*{Case study}
We demonstrate and compare the methods by addressing the emerging variable selection problem studied in Hu et al. \cite{hu2021variable} using the SWAN  data. The SWAN  was a multicenter, longitudinal study in the U.S. with the goal  of understanding women’s health across the menopause transition. The SWAN study enrolled 3305 women aged between 42 and 52 in 1996-1997 and followed them up to 2018 annually. The emerging research question is the identification of important risk factors for 3-year incidence of metabolic syndrome. Metabolic syndrome is a cluster of conditions that occur together and has been shown to increase the risk of heart disease, stroke and type 2 diabetes \cite{kazlauskaite2020midlife}. Identifying important predictors of metabolic syndrome among middle-aged women would be valuable to forming preventive interventions and reducing risks or threats to women's health.  This important research question has been less studied in the literature for women during their middle years.    

We use the analysis dataset described in Hu et al. \cite{hu2021variable}, which 
included 2313 women who did not have metabolic syndrome at enrollment. Among the 2313 women, 251 (10.9\%) developed metabolic syndrome within three years of enrollment, 1240 (53.6\%) did not, and the remaining 822 (35.5\%) had missing outcomes. Sixty candidate predictors (29 continuous variables and 31 discrete variables) were selected based on the previous literature \cite{han2019dietary,kazlauskaite2020midlife,janssen2008menopause,feng2017low} on risk factors for metabolic syndrome, including  the demographics,  daily life behaviour, dietary habits, sleep habits, medications,  menopausal status and related factors and mental status among others.  A detailed list of names and definitions of the 60 candidate variables is provided in Supplementary Table 12. Among the 60 candidate predictors 49 had missing values, and the amount of missing data in these variables ranges from 0.1\% to 27.1\%. Only 763 (33.0\%) participants had fully observed data.

\section*{Results}
Throughout, we used $\code{missForest}$  \cite{stekhoven2012missforest} to impute missing data for all methods considered. $\code{missForest}$ is a 
flexible imputation algorithm,  which recasts the missing data problem as a prediction problem and uses the random forest \cite{breiman2001random} for prediction.  The missing values are imputed in turn for each variable from a fitted forest regressing that variable on all other variables \cite{tang2017random}. As suggested by Tang et al.\cite{tang2017random} and Hu et al. \cite{hu2021variable},  $\code{missForest}$ performs better than $\code{MICE}$ \cite{van1999multiple}  across various missing data settings. All variables (predictor and response variables) available to the analyst were included in the imputation model without variable selection or specification of functional forms. Details of the imputation appear in Supplementary Section 1.

\subsection*{Simulation results}
Table~\ref{tab:simres} summarizes the variable selection results of each method when performed on the fully observed  data, incomplete data and complete cases only, for $n=1000$, 40 noise predictors, and 30\% and 60\% overall missingness. It is apparent that the performance of bootstrap based methods on incomplete data depends on the threshold $\pi$,  the optimal value of which varies by methods. By comparison, the performance of RR-BART does not depend on any additional parameters. Even with a large proportion of missingness, RR-BART achieves good variable selection performances (on the bases of AUC, precision, recall, $F_1$ and Type I error) that are  similar to the best performances BI-BART or BI-XGB could achieve with the optimal threshold values of $\pi$. Both BART and XGBoost had substantially deteriorated performance when only complete cases were used. Both MIA based methods had subpar performances (only slightly better than those of the complete-case analyses), with the missing outcomes either imputed or excluded. 

Supplementary Table 3--6 summarize simulation results for different sample sizes $n=300, 650, 1000, 5000$ and for $\alpha = 0.01, 0.05, 0.1$. When the sample size is small ($n=300, 650$), no methods provided satisfactory performance; in the presence of missing data, the proposed method could still recover the performance achievable on fully observed data. The findings are in agreement with the previous literature \cite{hu2021variable}. When the sample size is large ($n=1000,5000$),  all three methods RR-BART, BI-BART and BI-XGB, delivered good performance, and comparatively better performance with a smaller missingness proportion and more noise predictors. When the sample size increased from $n=1000$ to $n=5000$, the performance gain was only moderate, suggesting that $n=1000$ is sufficiently large of a sample size for implementing the proposed method. Bigger values of $\alpha$ tend to lead more selected variables, thus increased Type I error and recall. As suggested by an anonymous reviewer, we also included a ``baseline'' method, referred to as RR-BART (median), by which predictors with the posterior mean VIP exceeding the median value of the VIPs are selected.  The performance from the baseline method is comparatively lower than the proposed RR-BART method.

Figure 1 further highlights that the AUCs of BI-BART and BI-XGB  depend on the selection threshold $\pi$, with the highest AUC achieved at $\pi=0.1$ for BI-BART and $\pi=0.3$ for BI-XGB. The proposed RR-BART boasts the same highest AUC without the need to specify a value for $\pi$. The performance curves of the other four  metrics are shown in Supplementary Figure 1, which convey the same message. 

A perusal of Figure 2 suggests that RR-BART has a substantially higher power for selecting a discrete variable of even moderate effect size ($X_2$) than BI-BART and BI-XGB; meanwhile, maintains comparable power for identifying the continuous variables, even in complex conditions of nonadditivity and nonlinearity. 

In the extreme case where all of 10 predictors were useful, the average minimum VIP score was around 0.08, which is closer to 1/10 than is to 0. Note that the VIPs of all predictors sum to one. Supplementary Table 7 presents the distribution of the minimums of posterior mean VIPs across 250 data replications. The simulation results of this extreme scenario are provided in Table 2 and Supplementary Table 8. By definition, for \emph{all} methods, the precision is one and the Type I error becomes not applicable. There was a substantial drop in the $F_1$ score for both BART and XGBoost on fully observed data. This is because the comparatively less important variables were not selected, leading to a lower recall and in turn a lower $F_1$ score. Turning to the situations in which missing covariate and outcome data are present. When the missingness proportion is smaller (30\% overall missingness), the proposed RR-BART delivered similar performance, judged by $F_1$ and recall, as  BI-BART and BI-XGB. When the missingness proportion is higher (60\% overall missingness), RR-BART had a lower recall and $F_1$ score than the two bootstrap imputation based methods. Because the minimum VIP was considerably far away from 0, RR-BART step 2 deemed variable selection as unnecessary and thus selected all predictors. This modified algorithm, RR-BART (all selected), produced near-perfect performance. 

Supplementary Table 9  summarizes the numbers of times RR-BART performed better than, worse than or equal to BI-BART, across all simulation configurations. Overall, the two methods produced very similar performance. The small Monte Carlo errors of the variable VIP scores (see Supplementary Table 10) indicate that the numbers of simulation replications are sufficiently large. 

To assess whether the normality assumption underlying Rubin's rule is satisfied, Supplementary Figure 2 plots the posterior distributions of the VIPs for the 10 useful predictors, for the simulation scenario with $n=1000$, 40 noise predictors, and  60\% overall missingness proportion. It is reasonable to assume that the VIP scores are normally distributed. In addition, we also explored the strategy, recommended by Zou and Reiter \cite{zhou2010note} and Hu et al. \cite{hu2022flexible}, which obtains the posterior inferences for the variable VIPs by pooling posterior samples across model fits arising from the multiple datasets. Supplementary Table 11 shows that this strategy produced similar variable selection results as the proposed RR-BART  that uses Rubin's rule. 

\subsection*{Re-analysis of SWAN data}
Table~\ref{tab:case_results} lists the selected important predictors of 3-year incidence of metabolic syndrome by RR-BART, BI-BART $\pi = 0.1$ and BI-XGB $\pi = 0.3$, using the SWAN data. Note that we used the optimal threshold values of $\pi$ for BI-BART and BI-XGB that were suggested in our simulation study. Both RR-BART and BI-BART identified 17 predictors, among them 15 were the same;  11 predictors were selected by all three methods. In addition to the common risk factors for metabolic syndrome such as the diastolic blood pressure (DIABP), systolic blood pressure (SYSBP), lipoprotein(a) (LPA), and  triglycerides (TRIGRES), RR-BART was able to identify predictors such as tissue plasminogen activator (TPA) and apolipoprotein A-1 (APOARES) that were less studied in the literature for women's health. The selection of these two predictors has substantial justification, as levels of TPA antigen and low apolipoprotein A-1 (APOARES) were found to be associated with insulin resistance, which was involved in the pathogensis of impaired fasting glucose \cite{rao2004impaired, feng2017low}.

We further evaluated how well the selected variables predict the 3-year incidence of metabolic syndrome. A useful measure to assess the performance of risk prediction models is the validation plot \cite{steyerberg2010assessing}. A perfect risk prediction model would produce a $45^{\circ}$ line, with the predicted risk perfectly aligns with the observed event probability. Figure 3 compares the prediction performance of the BART model and the XGBoost model with the predictors selected by RR-BART, BI-BART and BI-XGB.  When implemented on the SWAN data, all three models performed well and produced similar AUCs. The validation lines show that RR-BART and BI-BART had better prediction performance for higher risk individuals.

\section*{Discussion}
We investigate strategies for variable selection with missing covariate and outcome data in large-scale health studies.  Prior work has shown that combining bootstrap imputation with tree-based variable selection methods has good operating characteristics with respect to selecting the most relevant predictors \cite{hu2021variable}. The computational cost of this method, however, can be high in large datasets, given that both bootstrap and recursive nonparametric modeling are needed. We propose an inference-based method that leverages the Bayesian machine learning model, BART, and uses Rubin's rule to combine the estimates and variances of the VIPs of each predictor provided by the BART models across multiply imputed datasets for variable selection.  In addition, we implement MIA within BART and XGB, which accommodate missing covariates by modifying the binary splitting rules. 

Our simulation study suggests that the proposed method RR-BART can achieve the best performance, with respect to AUC, $F_1$ score, precision, recall and type I error,  delivered by the BI-BART or BI-XGB method with the optimal selection threshold value of $\pi$.  RR-BART can reasonably well recover the performance a BART or XGBoost model can achieve on the fully  observed data,  in situations where the proportion of missingness is large and in complex conditions of nonlinearity and nonadditivity. Furthermore, it has been shown that tree-based methods tend to have a lower power in detecting a discrete predictor \cite{hu2021variable}. RR-BART has demonstrated a strong ability of identifying discrete variables, even those which have only moderate effect sizes. In general, using MIA with BART or XGB does not provide satisfactory performance of variable selection on incomplete data.  Moreover, RR-BART recognizes the situation that variable selection may not be needed if the minimum of average posterior mean VIPs  is considerably far away from zero.

The computational savings offered by RR-BART are substantial. All simulations were run in $\R$ on an iMAC with a 4 GHz Intel Core i7 processor. On a dataset of size $n=1000$  with 50 predictors, each RR-BART  took 4 minutes to run, while each BI-XGB implementation took 7 minutes and each BI-BART took about 55 minutes to run. More importantly, unlike the bootstrap imputation based methods,  RR-BART does not require a selection threshold $\pi$, whose optimal value varies by methods and simulation settings, and is not known \emph{a priori}. The performance of RR-BART does depend on the parameter $\alpha$, for which a default value of 0.05 is recommended.

Although our proposed method RR-BART provides promising performance for variable selection in the presence of MAR covariate and outcome data, it is possible to further improve the method for the big-$n$-small-$p$ situation, which is frequently encountered in health registry data. Another possible research avenue is to derive an alternative nonparametric measure of variable importance rather than the VIP that reflects  the impact of inclusion or deletion of predictors on the model prediction accuracy in the presence of missing data \cite{williamson2021nonparametric}. Finally, extending the methods to accommodate MNAR data could be a worthwhile contribution. 

\section*{Conclusion}
The proposed variable selection method, RR-BART, is both computationally efficient and practically useful in identifying important predictors using large-scale healthcare databases with missing data. 

\section*{Abbrevations}
 
 MAR: Missing at random \\
 MCAR: Missing completely at random  \\
 MNAR: Missing not at random \\
 BART: Bayesian additive regression trees \\
 XGBoost: Extreme gradient boosting \\
 BI-BART: Bootstrap imputation with Bayesian additive regression trees \\
 BI-XGB: Bootstrap imputation with extreme gradient boosting\\
 RR-BART: Rubin's rule implemented with  Bayesian additive regression trees\\
 AUC: Area under the curve \\ 
 MCMC: Markov Chain Monte Carlo \\
 VIP: Variable inclusion proportion \\
 MIA: Missingness incorporated in attributes \\
 SWAN: Study of Women's Health Across the Nation \\
 DIABP: Diastolic blood pressure \\
 SYSBP: Systolic blood pressure \\
 LPA: Lipoprotein(a) \\
 TRIGRES: Triglycerides \\
 TPA: Plasminogen activator \\
 APOARES: Apolipoprotein A-1 \\
 
\section*{Ethics declarations} 

\begin{backmatter}
\section*{Ethics approval and consent to participate}
All methods were performed in accordance with the relevant guidelines and regulations. All methods implementation, simulations and statistical analyses were performed using $\R$ Statistical Software (version 4.0.4; R Foundation for Statistical Computing, Vienna, Austria). Our study is not human subjects research as it consists of only simulated datasets and de-identified data. Because the SWAN data used in our case study are completely de-identified, our institution’s institutional review board does not require an approval for research involving SWAN data. 

\section*{Consent for publication}
Not applicable. 

\section*{Competing interests}
  The authors declare that they have no competing interests.
  
 \section*{Availability of data and materials}
Codes for implementing all methods and replicating simulations are available in the GitHub page of the corresponding author \url{https://github.com/liangyuanhu/RR-BART}. Deidentified SWAN data are publicly available at \url{https://www.swanstudy.org/swan-research/data-access}.    

 \end{backmatter}
 
 \section*{Funding}
 This work was supported in part  by award ME\_2017C3\_9041 from the Patient-Centered Outcomes Research Institute, and by grants R21CA245855 and P30CA196521-01 from the National Cancer Institute. 
 
\section*{Author's contributions}
   All authors have read and approved the manuscript. JL wrote the final version of the statistical codes, drafted the first version of the manuscript, and reviewed the final version of the paper. LH developed the methods and conceived the simulation study, assembled the team,  supervised coding, and finalized the manuscript.  CH worked on extracting and cleaning the SWAN data, wrote the first version of the statistical codes and reviewed the final version of the paper. JJ validated the final version of the codes for simulations and data analyses, and reviewed the final version of the paper. SL worked on interpreting the results and on finalizing the manuscript. UG worked on interpreting the results, deciding on its presentation and reviewed  the final version of the manuscript.

\section*{Acknowledgements}
Not available.


\bibliographystyle{bmc-mathphys} 
\bibliography{references}      




\section*{Figures}
  \begin{figure}[h!]
  \caption{\csentence{The mean cross-validated AUC,  averaged across 250 data replications, for each of three methods: RR-BART, BI-BART and BI-XGB. The mean AUC for bootstrap imputation based methods BI-BART and BI-XGB varies by the threshold value of $\pi$. $\code{missForest}$ was used for imputation. The sample size $n=1000$. The proportion of missingness is 40\% in the outcome $Y$ and is 60\% overall. }
     }
      \end{figure}

\begin{figure}[h!]
  \caption{\csentence{Power of each of three methods, RR-BART, BI-BART and BI-XGB, for selecting each of 10 useful predictors across 250 data replications.  $\code{missForest}$ was used for imputation. The sample size $n=1000$. The proportion of missingness is 40\% in the outcome $Y$ and is 60\% overall.}
      }
      \end{figure}
      
       \begin{figure}[h!]
  \caption{\csentence{Validation plot of predicted probabilities of 3-year incidence of metabolic syndrome among middle-aged women using the SWAN data. The risk prediction models were the BART and XGBoost models with predictors selected via RR-BART, BI-BART and BI-XGB. $\code{missForest}$ was used for imputation. }
      }
     \end{figure}


\section*{Tables}
\setlength{\tabcolsep}{12pt} 
\begin{table}[htbp]
    \centering
    \caption{Simulation results for each variable selection approach performed on the fully observed data and among incomplete data. For bootstrap imputation based methods on incomplete data, we show results corresponding to different threshold values of $\pi$. The optimal value of $\pi$ leading to the highest $F_1$ score is $\pi = 0.1$ for BI-BART and $\pi = 0.3$ for BI-XGB. The sample size is $n=1000$. The number of useful predictors is 10 and the number of noise predictors is 40. Two missingness proportions were considered: 40\% missingness in $Y$ and 60\% overall missingness; 20\% missingness in $Y$ and 40\% overall missingness. The performance measures were computed across 250 data replications.  } 
    \bgroup
\def\arraystretch{1.1} 
    \begin{tabular}{lccccc}
    \toprule
        &AUC  &Precision &Recall &$F_1$  &Type I error \\\hline
      \multicolumn{6}{c}{\textbf{Fully observed data}}\\
        BART & 0.92 (0.88, 0.96) & 1.00 & 0.87 & 0.93 & 0.00\\
        XGBoost & 0.88 (0.84, 0.92) & 0.93 &0.81 & 0.86 & 0.02 \\
                \midrule
          \multicolumn{6}{c}{\textbf{Incomplete data: 40\% missingness in $Y$ and 60\% overall missingness}}\\
          RR-BART & 0.82 (0.78, 0.86) & 0.87 & 0.80 & 0.83 & 0.01\\
          BI-BART $\pi=0.1$ & 0.83 (0.78, 0.88) & 0.87 & 0.82 & 0.85 & 0.03 \\
          BI-BART $\pi=0.2$ & 0.81 (0.76, 0.86) & 0.97 & 0.71 & 0.82 & 0.01 \\
          BI-BART $\pi=0.3$ & 0.77 (0.72, 0.82) & 0.99 & 0.63 & 0.77 & 0.00 \\
          BI-BART $\pi=0.4$ & 0.73 (0.68, 0.78) & 1.00 & 0.55 & 0.71 & 0.00 \\
          BI-BART $\pi=0.5$ & 0.67 (0.62, 0.72) & 1.00 & 0.48 & 0.65 & 0.00 \\ 
          BI-BART $\pi=0.6$ & 0.59 (0.54, 0.64) & 1.00 & 0.40 & 0.57 & 0.00 \\ 
          BI-BART $\pi=0.7$ & 0.49 (0.44, 0.54) & 1.00 & 0.31 & 0.47 & 0.00 \\ 
          BI-BART $\pi=0.8$ & 0.41 (0.36, 0.46) & 1.00 & 0.23 & 0.38 & 0.00 \\ 
          BI-BART $\pi=0.9$ & 0.33 (0.28, 0.38) & 1.00 & 0.14 & 0.25 & 0.00 \\
          BI-BART $\pi=1.0$ & 0.25 (0.20, 0.30) & 1.00 & 0.07 & 0.13 & 0.00 \\
          BI-XGB $\pi=0.1$ & 0.58 (0.53, 0.63) & 0.40 & 0.88 & 0.55 & 0.36 \\ 
          BI-XGB $\pi=0.2$ & 0.74 (0.69, 0.79) & 0.66 & 0.85 & 0.75 & 0.15 \\
          BI-XGB $\pi=0.3$ & 0.82 (0.77, 0.87) & 0.83 & 0.83 & 0.83 & 0.03 \\
          BI-XGB $\pi=0.4$ & 0.76 (0.71, 0.81) & 0.96 & 0.72 & 0.82 & 0.01 \\
          BI-XGB $\pi=0.5$ & 0.70 (0.65, 0.75) & 0.99 & 0.63 & 0.77 & 0.00 \\
          BI-XGB $\pi=0.6$ & 0.64 (0.59, 0.69) & 1.00 & 0.54 & 0.70 & 0.00 \\
          BI-XGB $\pi=0.7$ & 0.59 (0.54, 0.64) & 1.00 & 0.41 & 0.58 & 0.00 \\
          BI-XGB $\pi=0.8$ & 0.47 (0.42, 0.52) & 1.00 & 0.29 & 0.44 & 0.00 \\
          BI-XGB $\pi=0.9$ & 0.35 (0.30, 0.40) & 1.00 & 0.17 & 0.29 & 0.00 \\
          BI-XGB $\pi=1.0$ & 0.28 (0.23, 0.33) & 1.00 & 0.04 & 0.09 & 0.00 \\
          MIA-BART (Impute missing Y) & 0.75 (0.71, 0.79) & 0.80 & 0.75 & 0.77 & 0.04\\
          MIA-BART (Exclude missing Y) & 0.72 (0.66, 0.78) & 0.78 & 0.70 & 0.74 & 0.05\\
          MIA-XGB (Impute missing Y) & 0.74 (0.70, 0.78) & 0.81 & 0.73 & 0.77 & 0.04\\
          MIA-XGB (Exclude missing Y) & 0.71 (0.65, 0.77) & 0.75 & 0.71 & 0.73 & 0.08\\

        BART Complete cases  & 0.70 (0.63, 0.77) & 0.90 & 0.60 & 0.72 & 0.03\\
        XGBoost Complete cases & 0.73 (0.66, 0.80) & 0.90 & 0.68 & 0.77 & 0.04 \\
        \midrule
        \multicolumn{6}{c}{\textbf{Incomplete data: 20\% missingness in $Y$ and 30\% overall missingness}}\\
         RR-BART & 0.86 (0.82, 0.90) & 0.91 & 0.84 & 0.87 & 0.02\\
         BI-BART $\pi=0.1$ & 0.87 (0.83, 0.91) & 0.91 & 0.87 & 0.89 & 0.01 \\
          BI-BART $\pi=0.2$ & 0.85 (0.80, 0.90) & 0.99 & 0.76 & 0.85 & 0.02 \\
          BI-BART $\pi=0.3$ & 0.82 (0.77, 0.87) & 1.00 & 0.68 & 0.82 & 0.01 \\
          BI-BART $\pi=0.4$ & 0.77 (0.72, 0.82) & 1.00 & 0.59 & 0.76 & 0.01 \\
          BI-BART $\pi=0.5$ & 0.72 (0.67, 0.77) & 1.00 & 0.54 & 0.70 & 0.00 \\ 
          BI-BART $\pi=0.6$ & 0.63 (0.58, 0.68) & 1.00 & 0.46 & 0.61 & 0.01 \\ 
          BI-BART $\pi=0.7$ & 0.53 (0.48, 0.58) & 1.00 & 0.36 & 0.51 & 0.00 \\ 
          BI-BART $\pi=0.8$ & 0.44 (0.39, 0.49) & 1.00 & 0.28 & 0.42 & 0.00 \\ 
          BI-BART $\pi=0.9$ & 0.38 (0.33, 0.43) & 1.00 & 0.18 & 0.30 & 0.00 \\
          BI-BART $\pi=1.0$ & 0.29 (0.24, 0.34) & 1.00 & 0.12 & 0.18 & 0.00 \\
          BI-XGB $\pi=0.1$ & 0.60 (0.55, 0.65) & 0.44 & 0.90 & 0.58 & 0.30 \\ 
          BI-XGB $\pi=0.2$ & 0.76 (0.71, 0.81) & 0.69 & 0.87 & 0.77 & 0.11 \\
          BI-XGB $\pi=0.3$ & 0.84 (0.79, 0.89) & 0.86 & 0.85 & 0.85 & 0.02 \\
          BI-XGB $\pi=0.4$ & 0.78 (0.73, 0.83) & 0.99 & 0.75 & 0.84 & 0.01 \\
          BI-XGB $\pi=0.5$ & 0.73 (0.68, 0.78) & 1.00 & 0.65 & 0.79 & 0.00 \\
          BI-XGB $\pi=0.6$ & 0.67 (0.62, 0.72) & 1.00 & 0.57 & 0.73 & 0.00 \\
          BI-XGB $\pi=0.7$ & 0.62 (0.57, 0.67) & 1.00 & 0.44 & 0.61 & 0.00 \\
          BI-XGB $\pi=0.8$ & 0.49 (0.44, 0.54) & 1.00 & 0.32 & 0.47 & 0.00 \\
          BI-XGB $\pi=0.9$ & 0.38 (0.32, 0.42) & 1.00 & 0.20 & 0.33 & 0.00 \\
          BI-XGB $\pi=1.0$ & 0.31 (0.25, 0.35) & 1.00 & 0.08 & 0.14 & 0.00 \\
          MIA-BART (Impute missing Y) & 0.78 (0.74, 0.82) & 0.83 & 0.77 & 0.79 & 0.03\\
          MIA-BART (Exclude missing Y) & 0.76 (0.70, 0.82) & 0.81 & 0.74 & 0.78 & 0.04\\
          MIA-XGB (Impute missing Y) & 0.77 (0.73, 0.82) & 0.83 & 0.76 & 0.79 & 0.03\\
          MIA-XGB (Exclude missing Y) & 0.73 (0.67, 0.79) & 0.78 & 0.73 & 0.75 & 0.06\\
          BART Complete cases  & 0.73 (0.67, 0.79) & 0.92 & 0.64 & 0.75 & 0.03\\
        XGBoost Complete cases & 0.76 (0.70, 0.82) & 0.93 & 0.71 & 0.80 & 0.03 \\
        \bottomrule
  \end{tabular}     
    \egroup  
    \label{tab:simres}
\end{table}

\begin{table}[htbp]
    \centering
    \caption{Simulation results for the setting in which there are 10 useful predictors and no noise variables. For bootstrap imputation methods on incomplete data, we show results corresponding to the best threshold values of $\pi$ based on $F_1$. The sample size $n=1000$. Two missingness proportions were considered: 40\% missingness in $Y$ and 60\% overall missingness; 20\% missingness in $Y$ and 40\% overall missingness. The performance measures were computed across 250 data replications. } 
    \bgroup
\def\arraystretch{1.1} 
    \begin{tabular}{lccccc}
    \toprule
        &AUC  &Precision &Recall &$F_1$  &Type I error \\\hline
      \multicolumn{6}{c}{\textbf{Fully observed data}}\\
        BART & 0.74 (0.68, 0.80) & 1.00 & 0.62 & 0.70 & NA\\
        XGBoost & 0.75 (0.69, 0.81) & 1.00 &0.61 & 0.69 & NA \\
                \midrule
          \multicolumn{6}{c}{\textbf{Incomplete data: 40\% missingness in $Y$ and 60\% overall missingness}}\\
          RR-BART & 0.73 (0.67, 0.79) & 1.00 & 0.36 & 0.48 & NA\\
          RR-BART (all selected) & 0.97 (0.95, 0.99) & 1.00 & 1.00 & 1.00 & NA\\
          BI-BART $\pi=0.1$ & 0.73 (0.67, 0.79) & 1.00 & 0.38 & 0.50 & NA \\

          BI-XGB $\pi=0.2$ & 0.79 (0.73, 0.85) & 1.00 & 0.54 & 0.64 & NA \\
        BART Complete cases  & 0.50 (0.43, 0.57) & 1.00 & 0.15 & 0.35 & NA\\
        XGBoost Complete cases & 0.55 (0.48, 0.62) & 1.00 & 0.18 & 0.38 & NA \\
         MIA-BART (Impute missing Y) & 0.66 (0.60, 0.72) & 1.00 & 0.31 & 0.42 & NA\\
          MIA-BART (Exclude missing Y) & 0.63 (0.56, 0.70) & 1.00 & 0.25 & 0.40 & NA\\
          MIA-XGB (Impute missing Y) & 0.73 (0.67, 0.69) & 1.00 & 0.50 & 0.59 & NA\\
          MIA-XGB (Exclude missing Y) & 0.70 (0.62, 0.77) & 1.00 & 0.46 & 0.55 & NA\\
 \midrule
          \multicolumn{6}{c}{\textbf{Incomplete data: 20\% missingness in $Y$ and 30\% overall missingness}}\\
          RR-BART & 0.77 (0.72, 0.82) & 1.00 & 0.51 & 0.67 & NA\\
           RR-BART (all selected) & 0.98 (0.96, 0.99) & 1.00 & 1.00 & 1.00 & NA\\
           BI-BART $\pi=0.1$ & 0.75 (0.70, 0.80) & 1.00 & 0.50 & 0.69 & NA \\
           BI-XGB $\pi=0.2$ & 0.80 (0.75, 0.85) & 1.00 & 0.52 & 0.70 & NA \\
           
            MIA-BART (Impute missing Y) & 0.70 (0.65, 0.75) & 1.00 & 0.46 & 0.60 & NA\\
          MIA-BART (Exclude missing Y) & 0.67 (0.61, 0.73) & 1.00 & 0.43 & 0.57 & NA\\
          MIA-XGB (Impute missing Y) & 0.70 (0.65, 0.75) & 1.00 & 0.46 & 0.64 & NA\\
          MIA-XGB (Exclude missing Y) & 0.67 (0.61, 0.73) & 1.00 & 0.42 & 0.61 & NA\\
          BART Complete cases  & 0.54 (0.49, 0.59) & 1.00 & 0.16 & 0.39 & NA\\
           XGBoost Complete cases & 0.56 (0.51, 0.61) & 1.00 & 0.20 & 0.40 & NA \\
        \bottomrule
  \end{tabular}     
    \egroup  
    \label{tab:extreme}
\end{table}


\begin{table}[htbp]
    \centering
    \caption{Variable selection results by each of 3 methods, with the best imputation method and threshold value of $\pi$ for BI-BART  and BI-XGB suggested in simulations. BART was used with $\code{missForest}$ and $\pi=0.1$, XGB with $\code{missForest}$ and $\pi=0.3$. Definitions of the variable names appear in Web Table 1. RR-BART and BI-BART both selected 17 variables, and BI-XGB selected 16 variables. } 
    \begin{tabular}{lccc}
    \toprule
        Variables & RR-BART &BI-BART & BI-XGB \\
        \midrule
         TRIGRES & Yes & Yes & Yes\\
         SYSBP & Yes & Yes & Yes\\
         LPA & Yes & Yes & Yes\\
         DIABP & Yes & Yes & Yes\\
         WAIST & Yes & Yes & Yes\\
         INSULIN & Yes & No & Yes\\
         LUCRES & Yes & No & No\\
         APOARES & Yes & Yes & Yes\\
         EDUCATION & Yes & No & Yes\\
         BP & Yes & Yes & No\\
         WHRATIO & Yes & Yes & Yes\\
         BMI & Yes & Yes & Yes\\
         RACE & Yes & Yes & No\\
         TPA & Yes & Yes & Yes\\
         DTTLIN & Yes & Yes & No\\
         SHBG & Yes & Yes & Yes\\
         RESTLES & Yes & Yes & No\\
         GLUCOSE & No & Yes & Yes\\
         T & No & Yes & No\\
         E2AVE & No & No & Yes\\
         PAI1 & No & No & Yes\\
         \bottomrule
    \end{tabular}
    \label{tab:case_results}
\end{table}

\newpage 

\section*{Additional Files}
\subsection*{Additional file 1 --- Web-based Supplementary Materials}
\noindent Supplementary Section 1. Random Forest based Imputation Algorithm - missForest. 
  
\noindent Supplementary Section 2. Simulation Setup. 

\noindent Supplementary Section 3. Additional tables (Supplementary Table 1--12) and figures (Supplementary Figure 1--2) for simulation study and case study. 
    

\end{document}